\documentclass{aa}  
\usepackage{graphicx}
\usepackage{txfonts}

\def\pro2{\textsc{\bfseries pro2}}
\def\lte2{\textsc{\bfseries lte2}}
\def\atoms2{\textsc{\bfseries atoms2}}
\def\setf2{\textsc{\bfseries setf2}}
\def\line1prof{\mbox{\textsc{\bfseries line1}\raisebox{-0.5ex}{\bfseries--}\textsc{\bfseries prof}}}

\def\etal{{et\,al.}\ }

\newcommand{\Teff}{$T\mathrm{\hspace*{-0.4ex}_{eff}}$\,}
\newcommand{\logg}{$\log\,g$\hspace*{0.5ex}}

\newcommand{\gppr}{\stackrel{>}{\scriptstyle \sim}}
\newcommand{\gappr}{\raisebox{-0.4ex}{$\gppr $}}

\begin{document}
\title{Discovery of photospheric argon in very hot central stars of planetary
  nebulae and white dwarfs\thanks{Based on observations made with the NASA-CNES-CSA Far
Ultraviolet Spectroscopic  Explorer. FUSE is operated for NASA by the Johns
Hopkins University under NASA contract NAS5-32985.}}

\author{K\@. Werner\inst{1}
   \and T\@. Rauch\inst{1}
   \and J.~W\@. Kruk\inst{2}}

\institute{Institut f\"ur Astronomie und Astrophysik, Universit\"at T\"ubingen, Sand 1, 72076 T\"ubingen, Germany
   \and Department of Physics and Astronomy, Johns Hopkins University, Baltimore, MD 21218, USA}

\offprints{K\@. Werner\\ \email{werner@astro.uni-tuebingen.de}}

\date{Received; accepted}

\authorrunning{K. Werner, T. Rauch, J.W. Kruk}
\titlerunning{Discovery of argon in hot central stars and white dwarfs}

\abstract
%Context
{We report the first discovery of argon in hot evolved stars and white
  dwarfs. We have identified the \ion{Ar}{vii}~1063.55~\AA\ line in some of the
  hottest known (\Teff=\,95\,000$-$110\,000~K) central stars of planetary nebulae
  and (pre-) white dwarfs of various spectral type.}
%Aim
{We determine the argon abundance and compare it to theoretical predictions from
  stellar evolution theory as well as from diffusion calculations.}
%Methods
{We analyze high-resolution spectra taken with the \emph{Far Ultraviolet
  Spectroscopic Explorer}. We use non-LTE line-blanketed model atmospheres and
  perform line-formation calculations to compute synthetic argon line profiles.}
%Results
{We find a solar argon abundance in the H-rich central star NGC~1360 and in the
  H-deficient PG1159 star PG\,1424+535. This confirms stellar evolution modeling
  that predicts that the argon abundance remains almost unaffected by
  nucleosynthesis. For the DAO-type central star NGC~7293 and the hot DA white
  dwarfs PG\,0948+534 and RE\,J1738+669 we find argon abundances that are up to
  three 
  orders of magnitude smaller than predictions of calculations assuming
  equilibrium of radiative levitation and gravitational settling. For the hot DO
  white dwarf PG\,1034+001 the theoretical overprediction amounts to one dex.}
%Conclusions
{Our results confirm predictions from stellar nucleosynthesis calculations for
  the argon abundance in AGB stars. The argon abundance found in hot white
  dwarfs, however, is another drastic example that the current state of
  equilibrium theory for trace elements fails to explain the observations
  quantitatively.}


\maketitle
%
%________________________________________________________________

\begin{figure*}[tbp]
\includegraphics[width=0.85\textwidth]{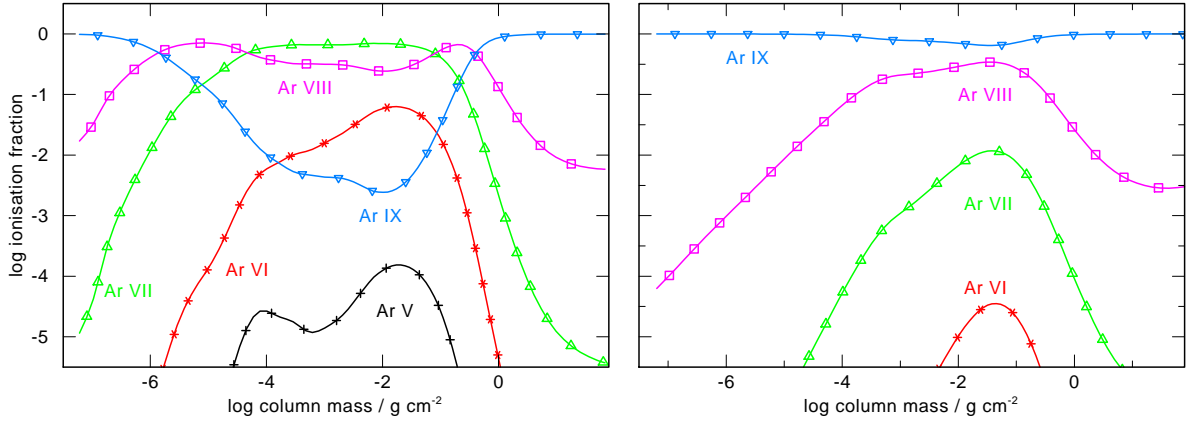}
  \caption[]{Ionization fraction of argon as a function of atmospheric
    depth. Left: Model with \Teff=\,110\,000~K, \logg=\,7.0 (for
    PG\,1424+535). Right: Model with \Teff=\,140\,000~K, \logg=\,7.0 (for
    PG\,1159-035). \ion{Ar}{vii} is dominant in the line-forming regions of the
    cooler model but is strongly depopulated in the hotter model, explaining the
    dependency of the \ion{Ar}{vii}~1063.55~\AA\ line strength from \Teff.
}
  \label{fig_ion}
\end{figure*}

\begin{figure*}[tbp]
\includegraphics[width=0.85\textwidth]{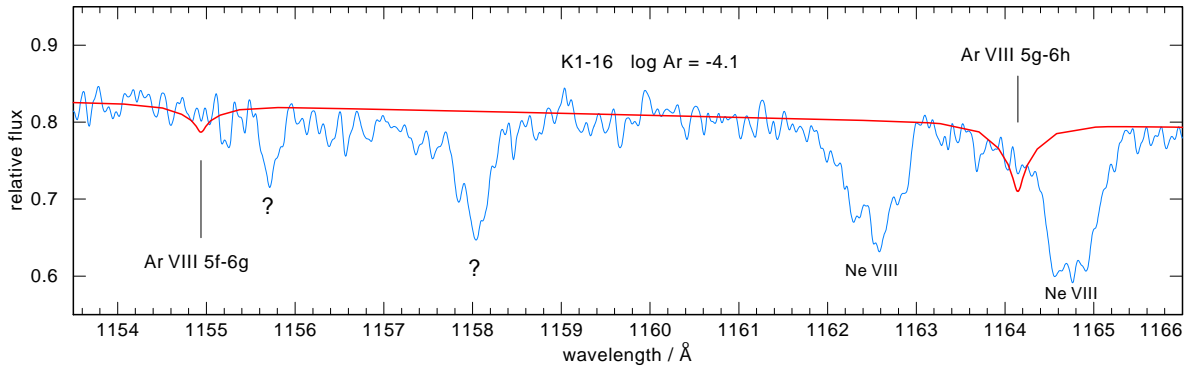}
  \caption[]{The strongest \ion{Ar}{viii} lines
    are exhibited by models with \Teff$\gappr$\,140\,000~K. They are located in a
    region where two unidentified and two \ion{Ne}{viii} photospheric lines are detected in the
    spectrum of K\,1-16. Unfortunately, the position of the
    \ion{Ar}{viii} lines is uncertain by $\pm 1.2$~\AA, preventing a unique
    identification. The parameters for the displayed model are those given for
    K\,1-16 in Tab.\,\ref{objects_tab} except for the Ar abundance. It is was set
    to $\log {\rm Ar}=-4.1$ (mass fraction).  }
  \label{fig_ar8}
\end{figure*}

\begin{figure}[tbp]
\includegraphics[width=0.95\columnwidth]{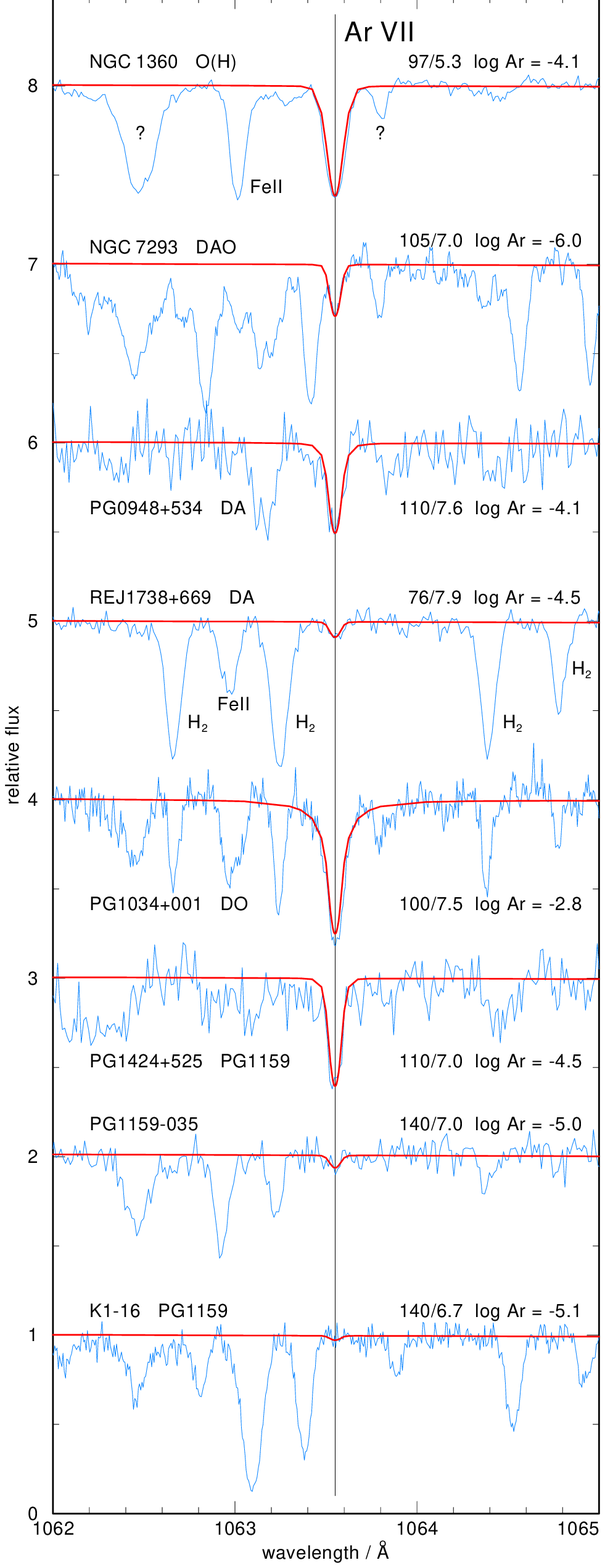}
  \caption[]{
Identification of the \ion{Ar}{vii}~1063.55~\AA\ line. It is not detectable in
the hottest objects (PG\,1159-035 and K\,1-16). Overplotted are synthetic line
profiles with $\log$~Ar (mass fraction), \Teff/kK, and \logg\ as labeled. Unidentified
photospheric lines at 1062.45 and 1063.8~\AA\ are seen in several objects
(marked by ``?'' in the top spectrum).  All other lines are from interstellar
\ion{Fe}{ii} and H$_2$; some of them are labeled.}
  \label{fig_all}
\end{figure}

\section{Introduction}
\label{intro}

The determination of elemental abundances in the extremely hot
hydrogen-deficient post-AGB stars of spectral type PG1159 is of particular
interest, because it is believed that these stars exhibit intershell matter as a
consequence of a late helium-shell flash. This allows one to directly study the
outcome of AGB star nucleosynthesis taking place in the region between the H-
and He-burning shells. While some elements are in line with expectations from
stellar evolution calculations, other elements clearly point at shortcomings in
the theory. For example, the extreme iron and sulfur depletion found in several
PG1159 stars is still not understood. On the other hand, the observed extreme
overabundances of fluorine confirms that AGB stellar modeling of complicated
nucleosynthesis processes is basically correct in other respects.

The detection of metals in PG1159 stars (and stars of other spectral types in
the same region of the HRD) is problematic because of their extremely high
effective temperatures (\Teff=\,75\,000$-$200\,000~K, Werner \& Herwig 2006). All
species are highly ionised so that spectral lines of elements heavier than C, N,
and O are, typically, from ionization stages \ion{}{v$-$viii}. The vast majority
of these transitions is at wavelengths below the hydrogen Lyman edge and, hence,
essentially inaccessible. However, a few exceptions are known and particularly
the \emph{Far Ultraviolet Spectroscopic Explorer} (FUSE) is playing a principal
role in the discovery of such highly ionised species. Two examples are the
\ion{Ne}{vii}~973.3~\AA\ and \ion{F}{vi}~1139.5~\AA\ lines which were never
identified before in any stellar atmosphere (Werner \etal 2004, 2005).

The argon abundance determination in stellar atmospheres is a challenge. Similarly
to neon, the solar abundance of this element cannot be determined directly
because of the lack of suitable spectral lines. The currently adopted solar
argon abundance (Table~\ref{objects_tab}) has been estimated from solar coronal
emission lines and solar energetic particles (Asplund \etal 2005). To our best
knowledge, the only stellar photospheric argon abundance determination was
performed for B stars which exhibit weak \ion{Ar}{ii} lines in the optical band
(the derived mean Ar abundance is 0.3~dex oversolar; Keenan \etal 1990, Holmgren
\etal 1990).

Taresch \etal (1997) were the first who considered argon in model atmosphere
analyses for even hotter stars. From their models tailored to the analysis of UV
and FUV spectra of the O3~If supergiant HD\,93129A (\Teff=\,52\,000~K, \logg=\,3.95)
they pointed out that the \ion{Ar}{vi}/\ion{Ar}{vii} ionisation balance could be
used as an additional independent temperature indicator. However, the blend of
potential Ar lines exhibited by their models with strong interstellar H$_2$
prevented the successful use of this tool in their case. To our knowledge, this
investigation of highly ionised argon was never addressed again in subsequent
spectroscopic work. It was this investigation that inspired us to look for argon
in hot post-AGB stars.

In this paper we report the discovery of an absorption line of highly ionised
argon, namely \ion{Ar}{vii}~1063.55~\AA, in a number of hot post-AGB stars and
white dwarfs with \Teff$\approx$\,100\,000~K. We describe in some detail the argon
model atom and the line identification in Sect.\,\ref{analysis}. We present the
abundance analysis in Sect.\,\ref{results} and conclude in
Sect.\,\ref{conclusions}.

\begin{table*}
\caption{List of objects and measured argon abundances. For the two DA white
  dwarfs \Teff\ and \logg\ are rather uncertain. We list the Ar abundances that
  we derive for the different sets of published \Teff/\logg\ values. In the last
  column we note the technique used to derive these sets (LTE or NLTE models, and
  fits to hydrogen Balmer or Lyman lines).
\label{objects_tab}}
\begin{tabular}{l l r c c c c l }
      \hline
      \hline
      \noalign{\smallskip}
object  & spectral type & \Teff & \logg & \multicolumn{3}{c}{\underline{\quad logarithmic argon abundance\quad}}  &reference for \Teff, \logg\\
        &               & [K]   & (cgs) & \underline{mass fraction}&      \multicolumn{2}{c}{\underline{\quad number ratio\quad}} &\\
        &               &       &       &      Ar       & Ar/H & Ar/He &                \\
      \noalign{\smallskip}
      \hline
      \noalign{\smallskip}
NGC 1360      &O(H) CSPN& 97\,000 & 5.3 &$-4.1$&$-5.6$&      & Traulsen \etal 2005 \\
NGC 7293      & DAO CSPN& 105\,000& 7.0 &$-6.0$&$-7.5$&      & Napiwotzki 1999 \\
      \noalign{\smallskip}
PG\,0948+534  & DA      & 110\,000& 7.6 &$-4.1$&$-5.7$&      & Barstow \etal 2003a, (NLTE Balmer) \\
              &         & 126\,000& 7.3 &$-3.5$&$-5.1$&      & Liebert \& Bergeron 1995, (LTE Balmer)\\
      \noalign{\smallskip}
RE\,J1738+669 & DA (CSPN?)& 76\,000 & 7.9 &$-4.5$&$-6.1$&    & Barstow \etal 2003b, (NLTE Lyman)\\
              &           & 67\,000 & 7.8 &$-3.7$&$-5.3$&    & Barstow \etal 2003a, (NLTE Balmer)\\
              &           & 95\,000 & 7.9 &$-5.5$&$-7.1$&    & Finley \etal 1997, (LTE Balmer)\\
      \noalign{\smallskip}
PG\,1034+001  & DO (CSPN?)& 100\,000& 7.5 &$-2.8$&   &$-3.8$ & Werner \etal 1995 \\
PG\,1424+535  & PG1159  & 110\,000& 7.0 &$-4.5$&     &       & Dreizler \& Heber 1998 \\
PG\,1159-035  & PG1159  & 140\,000& 7.0 &$\leq-5.0$ & &      & Werner \etal 1991 \\
K\,1-16       & PG1159 CSPN&140\,000&6.4&$\leq-5.1$ & &      & Kruk \& Werner 1998 \\
PG\,1144+005  & PG1159  & 150\,000& 6.5 &$\leq-4.1$ & &      & Werner \& Heber 1991 \\
PG\,1520+525  & PG1159 CSPN&150\,000&7.5&$\leq-4.1$ & &      & Dreizler \& Heber 1998 \\
      \noalign{\smallskip}
      \hline
      \noalign{\smallskip}
The Sun       &         &         &     &$-4.6$&$-5.8$&$-4.8$ & Asplund \etal 2005\\
      \noalign{\smallskip}
      \hline
     \end{tabular}
\end{table*}

\section{Argon abundance analysis}
\label{analysis}

FUSE observations and data reduction for most of our program stars were
described in our previous work as indicated in the Introduction.  Because of the
higher gravity of our program stars (2$-$4~dex) compared to the supergiant
HD\,93129A one can expect that the \ion{Ar}{vi}/\ion{Ar}{vii} lines are
strongest at effective temperatures significantly higher than 50\,000~K. This is
because the ionization balance is shifted to lower ionisation stages with
increasing gravity (and hence increasing gas pressure) which is compensated by
increasing temperature. Our models show that the \ion{Ar}{vii} line is strongest
at \Teff$\approx$\,100\,000~K. In Fig.\ref{fig_ion} we display the depth
dependency of the Ar ionization fractions in the atmospheric models of two
PG1159 stars with different effective temperatures (110\,000 and 140\,000~K). The
ionization balance in the hotter model is shifted away significantly to
ionization stages higher than \ion{Ar}{vii}.

We have identified the \ion{Ar}{vii}~1063.55~\AA\ line and performed an Ar
abundance determination in six stars (see Sect.~\ref{results}). The following
objects do not show this line: PG\,1159-035, K\,1-16, PG\,1144+005, and
PG\,1520+525. This allows us to derive a meaningful upper abundance limit for these
PG1159 stars.

In the spectra of the PG1159 stars NGC~246 and H1504+65 there is no H$_2$ blend
at the \ion{Ar}{vii}~1063.55~\AA\ line position, however, \Teff\ and \logg are
such that the argon line has disappeared. The following objects have a too
strong H$_2$ line blend: The PG1159 stars PG\,1707+427, RX\,J2117+3412,
HS\,2324+397, NGC~7094, Abell~43, Longmore~4,  the hot DO white dwarf
KPD\,0005+5106, the hot DA white dwarf PG\,0038+199, the hydrogen-rich central
stars NGC~6853, LSS~1362, NGC~1535, and the hot helium-rich stars (i.e.,
spectral type O(He)) K\,1-27, HS\,2209+8229, HS\,1522+6615, and LoTr~4.

We have designed an argon model atom for NLTE line-formation calculations. These
are performed using and keeping fixed the physical structure (temperature,
densities) of line-blanketed NLTE model atmospheres which are described in
detail in Werner \etal (2004). In short, they are plane-parallel and in
hydrostatic and radiative equilibrium. Model parameters and references to the
previous analyses are given in Tab.\,\ref{objects_tab}. The models are composed
of H, He, C, O, and Ne. For the H-rich central star NGC~1360 we calculated
models with a solar composition. We do not adopt the oversolar helium abundance
suggested by Traulsen \etal (2005) because we regard this result as
uncertain. For the DAO central star NGC~7293 the elemental abundances are solar,
too, except for helium which is reduced to ${\rm He/H}=0.03$ by number. For the DAs we
computed H-dominated models setting He and metal abundances to 1\% solar. For
the DO PG\,1034+001 we fixed the C and O abundances to the values derived by
Werner \etal (1995) and set the unknown Ne abundance to $\log({\rm Ne/He})=-8$ (by
number). For each star we performed calculations with different argon content in
order to estimate the observed abundance and the error of the analysis.

The model atom considers ionization stages \ion{Ar}{v$-$ix}, represented by 1,
15, 20, 41, and 1 NLTE levels, respectively, plus a number of LTE levels. In the
ions \ion{Ar}{vi$-$viii} we include 21, 36, and 199 line transitions,
respectively. Due to convergence problems which have their origin in a
particular \ion{Ar}{vi} EUV line and are not understood, the models for objects
with \Teff$\leq$\,110\,000~K (except for PG\,1034+001) do not include \ion{Ar}{vi}
lines. This is not expected to have an influence on the calculated
\ion{Ar}{vii-viii} line profiles. Atomic data were taken from the
NIST\footnote{http://physics.nist.gov/ PhysRefData/ASD/index.html} and Opacity
(Seaton \etal 1994) and IRON (Hummer \etal 1993) Projects databases
(TIPTOPbase\footnote{ http://vizier.u-strasbg.fr/topbase/}). For all lines we
assumed quadratic Stark broadening for the profile calculation, except for
\ion{Ar}{viii} lines for which linear Stark broadening is appropriate. Let us
comment in some detail on the spectral lines relevant for the present work.

The strategic \ion{Ar}{vii}~1063.55~\AA\ line is the singlet transition
3p\,$^{1}$P$^{\rm o}$--3p$^2$\,$^1$D. The same transition in the sulfur
isoelectronic configuration gives rise to the \ion{S}{v}~1501.76~\AA\ line that
is well known in stellar UV spectroscopy. The exact wavelength position of the
\ion{Ar}{vii} line coincides in the
Kelly\footnote{http://cfa-www.harvard.edu/amdata/ampdata/kelly/kelly.html}
(1987) and {\sc Chianti}\footnote{http://wwwsolar.nrl.navy.mil/chianti.html}
(Young \etal 2003) databases and is in accordance with the energy levels
provided by Bashkin \& Stoner (1975). The Kentucky
database\footnote{http://www.pa.uky.edu/$^\sim$peter/atomic/} gives
$\lambda=1063.63$~\AA\ which disagrees with the other data sources and with the
line position observed in our FUSE spectra. The $f$-value for this line is
reasonably well known. There is only a small difference in the value quoted by
Taresch \etal (1997) and the {\sc Chianti} database ($f=0.113$) and the value
given in the Kentucky database ($f=0.130$). For our calculations we have adopted
$f=0.113$. We have checked with model calculations for the presence of another
\ion{Ar}{vii} line listed in Taresch \etal (1997) at 982.0~\AA. It is the
strongest component of an intercombination multiplet ($\lambda\lambda~981.97,
999.47, 1066.50$~\AA) that arises from the same lower level as the 1063.55~\AA\
line but with an upper level in the triplet system. The $f$-value of the
982.0~\AA\ line, however, is a factor of $\approx$\,100 smaller than that of the
1063.55~\AA\ line and as a consequence is undetectable. A number of other
\ion{Ar}{vii} lines appear in the Taresch \etal (1997) list
($\lambda\lambda~1060.1, 1060.4, 1067.4, 1136.7$~\AA), for which no level
energies are publicly available. They are not detected in our program stars
probably because they arise from too highly excited levels.

Concerning the \ion{Ar}{vi} ion, Taresch \etal (1997) list three lines of the
intercombination resonance multiplet 3p$^2$\,P$^{\rm o}$--3p$^2$\,$^4$P
($\lambda\lambda~998.4, 1000.2, 1013.3$~\AA). The $f$-values of all multiplet
components, however, are $6.6\cdot 10^{-5}$ and smaller. The resulting lines are
undetectable in our model spectra as well as in the FUSE spectra of our
stars. Four other lines listed ($\lambda\lambda~1008.8, 1013.3, 1164.5,
1169.4$~\AA) are very highly excited and not detectable in our program stars.

A look at Fig.\,\ref{fig_ion} suggests that stars with \Teff$\gappr 140\,000$~K
could exhibit lines even from \ion{Ar}{viii}. This is a hydrogen-like ion having
line transitions between levels with principal quantum numbers
n=5$\longrightarrow$6 in the FUSE spectral range. The two lines with the largest
$gf$-values are 5f$\longrightarrow$6g and 5g$\longrightarrow$6h located at
1154.9~\AA\ and 1164.1~\AA, respectively. The latter is stronger and it
is indeed clearly seen in the hot star models although it is strongly broadened
by linear Stark effect (Fig.\,\ref{fig_ar8}). Unfortunately the level energies
are not known accurately enough so that the line position is uncertain within
$\pm 1.2$~\AA\ (Kentucky database). Within that uncertainty range of the
5f$\longrightarrow$6g line a hitherto unidentified line at 1155.7~\AA\ can be
seen in the spectra of the hottest PG1159 stars, among them is K\,1-16
(Fig.\,\ref{fig_ar8}), however, the identification with \ion{Ar}{viii} is
unclear. The 5g$\longrightarrow$6h line is located in the blue wing of a strong
and broad \ion{Ne}{viii} line at 1164.7~\AA\ and cannot be identified
clearly. The use of these \ion{Ar}{viii} lines as a spectral diagnostic is
questionable in any case because for some levels that probably are important for
a proper NLTE model atom the energies are completely lacking.

The two \ion{Ne}{viii} lines shown in Fig.\,\ref{fig_ar8} are new
identifications. Exploratory modeling suggests that they are of photospheric
origin. We are preparing detailed NLTE line formation calculations and will
present results in future work.

\section{Results}
\label{results}

We present profile fits to the \ion{Ar}{vii}~1063.55~\AA\ line in
Fig.\,\ref{fig_all}. The results are given in Table~\ref{objects_tab}.  In the
following we comment on the individual objects in their order of appearance in
Fig.\,\ref{fig_all} and Table~\ref{objects_tab}.

For the white dwarfs we compare our results to the theoretical argon abundance
predictions of Chayer \etal (1995a,b). They are based on the assumption that
trace elements in H or He dominated white dwarf atmospheres are kept from
gravitational settling by radiative levitation. These authors present detailed
predictions for the Ar abundance in DA and DO white dwarfs as a function of
effective temperature and surface gravity.

\paragraph{NGC~1360} The central star of this planetary nebula is of spectral
type O(H), i.e., its optical spectrum is dominated by the hydrogen Balmer
lines. Hoare \etal (1996) derived \Teff=\,110\,000~K and \logg=\,5.6. A more recent
analysis using metal lines in UV spectra indicates slightly different values,
namely \Teff=\,97\,000~K and \logg=\,5.3 (Traulsen \etal 2005).  We find a marginal
Ar overabundance, $\log({\rm Ar/H})=-5.6$ (by number), that is within the error
limit identical to the solar abundance ($-5.8$).

\paragraph{NGC~7293} This planetary nebula nucleus is classified as a DAO white
dwarf, i.e., it is hydrogen-rich with an admixture of helium. The atmospheric
parameters were derived by Napiwotzki (1999) using optical spectra:
\Teff=\,105\,000~K, \logg=\,7.0, ${\rm He/H}=0.03$. With $\log({\rm Ar/H})=-7.5$ (by number)
we find a strong Ar \emph{underabundance} of a factor of $\approx 50$. We can
compare this result with the prediction of levitation theory. From Fig.\,2 in
Chayer \etal (1995b) we read $\log({\rm Ar/H})=-3.5$ for DA white dwarfs, i.e.,
an  extreme \emph{overabundance} (factor $\approx 200$). The fact that the star
has some residual helium in the atmosphere is probably unimportant for this
prediction (note that for DO white dwarfs the predicted Ar abundance is even
higher, by about 1~dex). In essence, the predicted Ar abundance is four orders
of magnitude higher than the observed one. Traulsen \etal (2005) have claimed
higher \Teff\ and lower \logg\ from UV spectra (120\,000~K, 6.5) which both have
the tendency to weaken the \ion{Ar}{vii} line and, hence, a 0.5~dex higher Ar
abundance is derived in this case which is still much below the theoretical
prediction.

\paragraph{PG\,0948+534} This is one of the hottest known DA white dwarfs. From
Balmer line NLTE fits Barstow \etal (2003a) derived \Teff=\,110\,000~K, \logg=\,7.6. Our
best fit model has $\log({\rm Ar/H})=-5.7$ (by number). The Chayer \etal (1995b)
prediction for these parameters is much higher, $\log({\rm Ar/H})=-3.9$. (A
slight extrapolation was needed because their calculations do not consider
\Teff$>$\,100\,000~K.) In their LTE Balmer line analysis Liebert \& Bergeron (1995) find \Teff\
even higher and \logg\ lower (\Teff=\,126\,000~K, \logg=\,7.3) which both has the
tendency to make the \ion{Ar}{vii} line weaker. We have computed a small model
set with their \logg\ and \Teff=\,130\,000~K and different Ar abundances. The best
fit is now achieved at $\log({\rm Ar/H})=-5.1$ which makes the difference with the
predicted abundance smaller but it is still significant.

\paragraph{RE\,J1738+669} This is another extremely hot DA white dwarf. The
Balmer line LTE analysis by Finley \etal (1997) resulted in \Teff=\,95\,000~K,
\logg=\,7.9. We derive $\log({\rm Ar/H})=-7.1$ (by number). The respective Chayer
\etal (1995b) prediction is $\log({\rm Ar/H})=-4.3$ which is again a significant
overestimation. The \Teff/\logg\ values for this DA are, however, rather
uncertain. Barstow \etal (2003a,b) have performed a separate analysis of the
Balmer and Lyman lines and arrived at significantly lower \Teff\ but with
discrepant values: 67\,000/7.8 and 76\,000/7.9, respectively. For these two
parameter sets we derive $\log({\rm Ar/H})=-5.3$ and $\log({\rm Ar/H})=-6.1$
which can be compared to the predicted diffusion abundances for the respective
\Teff/\logg\ sets: $-4.8$ and $-4.5$. This means an overestimation of 0.5 and
1.5 dex. Thus, irrespective of the uncertain temperature we find that theory
overpredicts the Ar abundance. Only if one accepts the lowest \Teff\ value then
the Ar abundance is close to the predicted amount.

\paragraph{PG\,1034+001} This DO is among the hottest known helium-rich white
dwarfs (\Teff=\,100\,000~K, \logg=\,7.5). It exhibits the strongest \ion{Ar}{vii}
line in all stars we examined. We derive $\log({\rm Ar/He})=-3.8$ (by
number). From Fig.\,16 in Chayer \etal (1995a) we find that the theoretical
prediction is about one dex higher: $\log({\rm Ar/He})=-2.9$.

\paragraph{PG\,1424+535} This is the only PG1159 star in which we could identify
the \ion{Ar}{vii} line. It is rather strong and we derive a mass fraction of
$\log {\rm Ar}=-4.5$. This is close to the solar value ($-4.6$).

\paragraph{PG\,1159-035} The prototype of the PG1159 class does not show clear
evidence for the \ion{Ar}{vii} line. We derive an upper limit of $\log {\rm
Ar}=-4.5$ (mass fraction) which is almost equal to the solar abundance value
($-4.6$).

\paragraph{K\,1-16, PG\,1144+005, PG\,1520+525} These PG1159 stars have
effective temperatures similar to the prototype. Hence the non-detection of
argon allows to derive the solar abundance value as an approximate upper limit,
too.

From our experience with the profile fitting we estimate the error of our
abundance determinations to 0.5~dex.

\section{Conclusions}
\label{conclusions}

\subsection{PG1159 stars and the H-rich central star NGC~1360}

Nucleosynthesis calculations for AGB stars predict that the argon abundance in
the intershell region remains almost unchanged. In a particular model
calculation for a 2~M$_\odot$ AGB star the most abundant argon isotopes
$^{36}$Ar and $^{38}$Ar are being destroyed and produced by neutron captures,
respectively, resulting in a net reduction of the total Ar abundance of 0.2~dex
after 30 thermal pulses (Gallino priv. comm.). This is in line with results from
argon isotope analyses of meteoritic SiC grains which were formed in the winds
from AGB carbon stars (Gallino \etal 1990). All five noble gases detected in
such grains have elemental and isotopic abundances which are very similar to
results of nucleosynthesis calculations for the intershell of these stars (Lewis
\etal 1990).

The solar abundance of argon that we detected in the PG1159 star PG\,1424+535
and in the H-rich central star NGC~1360 further confirm these
results. PG\,1424+535 displays intershell matter and proves that the Ar
abundance remained unchanged during the previous AGB evolution. NGC~1360 could
show deviations from the solar Ar abundance if third dredge-up of strongly
Ar-depleted/enhanced matter during the previous AGB evolution would have
occurred. This is not observed. The solar upper Ar abundance limits determined
for the three other PG1159 stars in our sample rule out an extraordinary argon
production.

\subsection{DA, DAO, and DO white dwarfs}

Chayer \etal (1995a,b) have presented detailed predictions about trace element
abundances based on diffusion/levitation equilibrium theory. They found,
however, that equilibrium theory is unable to reproduce the observed trace
element abundances quantitatively for most species. This failure has become
increasingly obvious during the last years when more and more observational
results became available.

Vennes \etal (2005) have for the first time discovered trans-iron group elements
in hot DAs (germanium and probably tin). The derived abundances are nearly solar
(i.e., of the order $10^{-8}-10^{-9}$ relative to H, by number) and this made
these authors ask: ``With the detection of such an unlikely candidate [Ge], what
of abundant elements such as Ne, Mg, Ar, and Ca?  These elements would be
detectable in high-dispersion extreme-ultraviolet spectra. Until their
abundances have been determined, the abundance pattern in hot DA white dwarfs
remains incomplete.''  Our results on argon are one step towards this goal. They
also show that far-ultraviolet data supplied by FUSE offer the only way to this
goal as long as high-resolution EUV spectroscopy is unavailable. A significant
number of still unidentified photospheric lines in the FUSE spectra of hot white
dwarfs (some examples are presented in Figs.\,\ref{fig_ar8} and \ref{fig_all})
may hold the key to even more hitherto unidentified elements in white
dwarfs. The argon abundances derived in the present paper are another example
that equilibrium theory is unable to reproduce the observed trace element
abundances quantitatively.

To conclude, we recall that not only PG1159 photospheric metal abundances are
affected by s-process nucleosynthesis that took place during the AGB phase. This
may also be the case for white dwarfs although diffusion has certainly altered
the abundance pattern. This is suggested by the recent discovery of strongly
oversolar abundances of trans-iron group elements (up to iodine) in cool
(\Teff$\approx$\,50\,000~K) DO white dwarfs (Chayer \etal 2005). In contrast, the
one dex oversolar argon abundance that we found in the hot DO
PG\,1034+001 cannot be the result of nucleosynthesis but must be the result of
selective radiative levitation.
 
\begin{acknowledgements}
We would like to thank Roberto Gallino for sending his results on
nucleosynthesis calculations, and Maria Lugaro for pointing out the argon
analysis of meteoritic SiC grains. T.R. is supported by the German Ministry of
Economy and Technology through DESY under grant 05\,AC6VTB. J.W.K. is supported
by the FUSE project, funded by NASA contract NAS5-32985.
\end{acknowledgements}

\end{document}